\documentclass[12pt,a4paper]{article}
\usepackage{a4wide}
\usepackage{epsf}
\usepackage{graphicx}
 \newcommand{\vphi}{\varphi}

\begin{document}

\title{New Black Hole Solutions with Axial Symmetry
in Einstein-Yang-Mills Theory}

\vspace{1.5truecm}
\author{
{\bf Rustam Ibadov}\\
Department of Theoretical Physics and Computer Science,\\
Samarkand State University, Samarkand, Usbekistan\\
and\\
{\bf Burkhard Kleihaus, Jutta Kunz and Marion Wirschins}\\
Institut f\"ur Physik, Universit\"at Oldenburg, Postfach 2503\\
D-26111 Oldenburg, Germany}

\vspace{1.5truecm}

\maketitle
\vspace{1.0truecm}

\begin{abstract}
We construct new black hole solutions in Einstein-Yang-Mills theory.
They are static, axially symmetric and asymptotically flat.  They are 
characterized by their horizon radius and a pair of integers $(k,n)$,
where $k$ is related to the polar angle and $n$ to the azimuthal angle.
The known spherically and axially symmetric EYM black holes have $k=1$.
For $k>1$, pairs of new black hole solutions appear above a minimal value 
of $n$, that increases with $k$.  Emerging from globally regular solutions,
they form two branches, which merge and end at a maximal value of the 
horizon radius. The difference of their mass and their horizon mass
equals the mass of the corresponding regular solution, as expected from 
the isolated horizon framework.

\end{abstract}
\vfill\eject

\section{Introduction}

The well-known regular Bartnik-McKinnon (BM) solutions \cite{bm}
and the corresponding non-Abelian black hole solutions \cite{bh},
are asymptotically flat, static spherically symmetric solutions
of SU(2) Einstein-Yang-Mills (EYM) theory.
They are unstable solutions, sphalerons \cite{stab}, and are
characterized by the number of nodes of the gauge field.
Besides these spherically symmetric solutions there are also 
asymptotically flat, static regular and black hole solutions,
which possess only axial symmetry \cite{kk}.
These are characterized by two integers,
the node number of their gauge field function(s),
and the winding number with respect to the azimuthal angle, denoted $n$.
The spherically symmetric solutions have winding number $n=1$,
while the only axially symmetric solutions have winding number $n>1$.
The $n>1$ black hole solutions possess an event horizon with
a slight elongation along the symmetry axis \cite{kk}.
All these EYM black hole solutions exist for arbitrarily large horizon size.

In SU(2) Einstein-Yang-Mills-Higgs (EYMH) theory with a triplet Higgs field,
gravitating monopole solutions and black holes with monopole hair arise 
\cite{mono,hkk}.
As in EYM theory, the regular EYMH solutions are characterized by two integers,
the node number of the gauge field and the azimuthal winding number $n$,
which corresponds to the topological charge of the monopoles.
The black hole solutions are characterized in addition by their horizon size.
These EYMH black hole solutions exist only up to a maximal value 
of the horizon size \cite{mono,hkk}, unlike the known EYM black hole solutions.

EYMH theory allows for further static axially symmetric solutions,
representing gravitating monopole-antimonopole pair, chain and vortex
solutions \cite{map,kks,kks4},
which can be characterized by the azimuthal winding number $n$,
and by a second integer $m$, related to the polar angle.
For the monopole-antimonopole chains, 
which arise in flat space for $n=1$ and 2, 
the integer $m$ corresponds to the number of nodes of the Higgs field
(and thus the number of poles on the symmetry axis),
while in vortex solutions, which arise in flat space for winding number $n>2$,
the Higgs field vanishes (for even $m$)
on $m/2$ rings centered around the symmetry axis \cite{kks}.
To all these regular solutions associated black hole solutions should exist,
obtained so far only for a monopole-antimonopole pair \cite{dipole}.

In the limit of vanishing Higgs expectation value, EYMH solutions
approach (after rescaling) EYM solutions \cite{mono,map,kks4}.
Interestingly, when $n \ge 4$ and $m \ge 4$,
new regular EYM solutions appear as limiting solutions \cite{ikks,kks4}.
These static axially symmetric solutions have been 
characterized by the integers $(k,n)$, where $2k=m$.
They have been constructed numerically for $k=2$, $n \ge 4$, 
and $k=3$, $n \ge 6$ \cite{ikks}.
Unlike the $k=1$ EYM solutions, the $k>1$ solutions always appear in pairs.
In this letter we construct the corresponding EYM black hole solutions.
As expected, a branch of black hole solutions is associated with
each regular solution. 
Intriguingly, the two branches of black hole solutions from a pair of 
regular solutions merge and end at a maximal value of the horizon size.

In the isolated horizon framework \cite{ashtekar,sudar,iso2}, 
non-Abelian black hole solutions can be
interpreted as bound states of regular solutions and 
Schwarzschild black holes \cite{iso2}.
Furthermore, the isolated horizon framework yields a relation
for the mass of non-Abelian black hole solutions,
representing it as the sum of the mass of the regular solution
and the horizon mass
of the black hole solutions \cite{sudar}.
We here show that this relation is also valid for the new black holes.
In particular, the two regular solutions of a given $k$ and $n$
are connected via this mass formula.

In section II we present the EYM action, the axially
symmetric ansatz and the boundary conditions. 
In section III we address their asymptotic and horizon properties.
We present our numericl results in section IV,
and we give our conclusions in section V.

\section{Action and Ansatz}

We consider the SU(2) EYM action
\begin{equation}
S=\int \left ( \frac{R}{16\pi G} 
-\frac{1}{2} {\rm Tr} (F_{\mu\nu} F^{\mu\nu}) \right ) \sqrt{-g} d^4x
\ \label{action} \end{equation}
with Ricci scalar $R$,
field strength tensor
\begin{equation}
F_{\mu \nu} = 
\partial_\mu A_\nu -\partial_\nu A_\mu + i e \left[A_\mu , A_\nu \right] 
\ , \label{fmn} \end{equation}
gauge potential $ A_{\mu} = \tau^a A_\mu^a/2 $,
and gravitational and Yang-Mills coupling constants $G$ and $e$,
respectively.
Variation of the action (\ref{action}) with respect to the metric
$g^{\mu\nu}$ leads to the Einstein equations,
variation with respect to the gauge potential $A_\mu$ 
to the gauge field equations.

In isotropic coordinates
the static axially symmetric metric reads \cite{kk}
\begin{equation}
ds^2=
  - f dt^2 +  \frac{m}{f} d r^2 + \frac{m r^2}{f} d \theta^2 
           +  \frac{l r^2 \sin^2 \theta}{f} d\varphi^2
\ , \label{metric} \end{equation}
where the metric functions
$f$, $m$ and $l$ are functions of 
the coordinates $r$ and $\theta$, only.
The $z$-axis ($\theta=0$, $\pi$) represents the symmetry axis.
Regularity on the $z$-axis requires $m=l$ there.

For the gauge field we employ the ansatz \cite{kk,kks,kks4,ikks}
\begin{equation}
A_\mu dx^\mu =
\frac{1}{2er} \left[ \tau^n_\varphi 
 \left( H_1 dr + \left(1-H_2\right) r d\theta \right)
 -n \left( \tau^{n,k}_r H_3 + \tau^{n,k}_\theta H_4 \right)
  r \sin \theta d\phi \right]
\ . \label{gf1} \end{equation}
Here the symbols $\tau^{n,k}_r$, $\tau^{n,k}_\theta$ and $\tau^n_\varphi$
denote the dot products of the cartesian vector
of Pauli matrices, $\vec \tau = ( \tau_x, \tau_y, \tau_z) $,
with the spatial unit vectors
\begin{eqnarray}
\vec e_r^{\ n,k}      &=& 
(\sin k \theta \cos n \varphi, \sin k \theta \sin n \varphi, \cos k\theta)
\ , \nonumber \\
\vec e_\theta^{\ n,k} &=& 
(\cos k \theta \cos n \varphi, \cos k \theta \sin n \varphi,-\sin k \theta)
\ , \nonumber \\
\vec e_\varphi^{\ n}   &=& (-\sin n \varphi, \cos n \varphi,0) 
\ , \label{rtp} \end{eqnarray}
respectively.
The gauge field functions $H_i$, $i=1-4$, depend on
the coordinates $r$ and $\theta$, only.
For $k=n=1$ and $H_1=H_3=0$, $H_2=1-H_4=w(r)$
the spherically symmetric black hole solutions \cite{bh} are recovered,
while for $k=1$, $n>1$, one obtains 
the axially symmetric solutions of \cite{kk}.
The new black hole solutions reported here are obtained for $k>1$.
In the limit of vanishing horizon radius they converge pointwise to
the globally regular solutions with the same integers $k$ and $n$ 
\cite{ikks}.
The globally regular solutions are related to EYMH solutions with $m=2k$ 
in the limit of vanishing Higgs field \cite{ikks,kks4}.

The ansatz is form-invariant under the Abelian gauge transformation
\cite{kk}
\begin{equation}
 U= \exp \left({\frac{i}{2} \tau^n_\phi \Gamma(r,\theta)} \right)
\ .\label{gauge} \end{equation}
We fix the gauge by choosing the gauge condition \cite{kk,kks,kks4,ikks}
\begin{equation}
 r \partial_r H_1 - \partial_\theta H_2 = 0 
\ . \label{gc1} \end{equation}

To obtain asymptotically flat solutions
which are regular at the horizon 
and possess the proper symmetries,
we need to impose appropriate boundary conditions \cite{kk,kks,kks4,ikks}.
The horizon of the non-Abelian black hole solutions
resides at a surface of constant radial coordinate $r=r_{\rm H}$.
At the horizon we impose the boundary conditions
\begin{equation}
f=m=l=0 \ , \ \ \ 
H_1 = 0 \ , \ \ \ \partial_r H_2 = \partial_r H_3= \partial_r H_4 = 0  \ ,
\label{bch_metric} 
\end{equation}
at infinity we impose 
\begin{equation}
f = m = l = 1 \ , \ \ \
H_1 =H_3=0 \ , \ H_2 = 1 - 2k \ , \ 
H_4 =2 \sin(k\theta)/\sin\theta \ ,
\label{bc_inf} 
\end{equation}
and on the $z$-axis we impose
\begin{equation}
\partial_\theta f=\partial_\theta m=\partial_\theta l =0 \ , \ \ \
 H_1=H_3=0\ , \ \partial_\theta H_2=\partial_\theta H_4=0 \ .
\label{bc_sym} 
\end{equation}

\section{\bf Properties}

We introduce the dimensionless coordinate 
$x=\frac{e}{\sqrt{4\pi G}} r$ and horizon radius 
$x_{\rm H}=\frac{e}{\sqrt{4\pi G}} r_{\rm H}$.
Defining the mass $M$ of the black hole
solutions via the Komar integral,
the dimensionless mass $\mu = \frac{e G}{\sqrt{4 \pi G }} M$,
is determined by the derivative of the metric function $f$
at infinity 
\begin{equation}
\mu = \frac{1}{2} \lim_{x \rightarrow \infty} x^2 \partial_x  f
\ . \label{mass} \end{equation}

From the equations of motion it follows \cite{kk}, that
the Kretschmann scalar is finite at the horizon,
and that the surface gravity $\kappa$ \cite{wald},
\begin{equation}
\kappa^2=-(1/4)g^{tt}g^{ij}(\partial_i g_{tt})(\partial_j g_{tt})
\ , \label{sg} \end{equation}
is constant, as required by the zeroth law of black hole physics.
Expansion of the metric functions near the horizon in the form
$$ f(x,\theta) = \tilde{x}^2 f_2(\theta)
                +O\left(\tilde{x}^3\right) \ , \ \ \ 
m(x,\theta) = \tilde{x}^2 m_2(\theta)
                +O\left(\tilde{x}^3\right) \ , 
$$
where $\tilde{x}= (x/x_{\rm H} -1)$,
yields the dimensionless surface gravity
${\displaystyle \hat{\kappa} = 
 \frac{f_2(\theta)}{x_{\rm H}\sqrt{m_2(\theta)}}}$
related to $\kappa$ by $\kappa = \hat{\kappa}e/\sqrt{4 \pi G}$.

We introduce the area parameter $x_\Delta$ \cite{ashtekar,sudar},
defined via the dimensionless area of the black hole horizon $A$,
\begin{equation}
A = 2 \pi x_{\rm H}^2
\int_0^\pi  d\theta \sin \theta
\left. \frac{\sqrt{l m}}{f} \right|_{x_{\rm H}^2} 
 = 4 \pi x_\Delta^2
\ . \label{area} \end{equation}

The deformation of the horizon is revealed, when
the circumference of the horizon along the equator, $L_e$,
is compared to the circumference of the horizon along the poles, $L_p$,
\begin{equation}
L_e = \int_0^{2 \pi} { d \vphi \left.
 \sqrt{ \frac{l}{f}} x \sin\theta
 \right|_{x=x_{\rm H}, \theta=\pi/2} } \ , \ \ \
L_p = 2 \int_0^{ \pi} { d \theta \left.
 \sqrt{ \frac{m  }{f}} x
 \right|_{x=x_{\rm H}, \vphi=const.} }
\ , \label{lelp} \end{equation}
since the black hole solutions have $L_p \ne L_e$ (in general).

The isolated horizon framework \cite{ashtekar,sudar}
yields an intriguing relation
between the ADM mass $\mu$ of a black hole with area parameter $x_\Delta$
and the mass $\mu_{\rm reg}$ of the corresponding globally regular solution
\cite{ashtekar},
\begin{equation}
\mu =  \mu_{\Delta}
+\mu_{\rm reg}
\label{ash} \ , \end{equation}
where the (dimensionless) horizon mass $\mu_{\Delta}$ is defined via 
\begin{equation}
\mu_{\Delta} = \int_0^{x_{\Delta}} \hat{\kappa}(x'_{\Delta}) 
 x'_{\Delta} d x'_{\Delta}
\label{Mhor} \ . \end{equation}

The isolated horizon formalism further suggests
to interpret a non-Abelian black hole as a bound
state of a regular solution and a Schwarzschild black hole \cite{iso2},
\begin{equation}
\mu = \mu_{\rm reg} + \mu_{\rm S} + \mu_{\rm bind} \ ,
\label{IHbs} \end{equation}
where $\mu_{\rm S} = x_\Delta/2$ is the ADM mass
of the Schwarzschild black hole with area parameter $x_\Delta$,
and $\mu_{\rm bind}$ represents the binding energy of the system,
\begin{equation}
\mu_{\rm bind}= \mu_\Delta-\mu_{\rm S} \ .
\label{IHbind} \end{equation}

\section{\bf Results}

Subject to the above boundary conditions,
we solve the system of seven coupled non-linear partial
differential equations numerically.
To map spatial infinity to the finite value $\bar{x}=1$,
we employ the radial coordinate 
\begin{equation}
\bar{x} = \frac{x-x_{\rm H}}{1+x} \ .
\   \label{barx} \end{equation}
The numerical calculations are based on the Newton-Raphson method,
and are performed with help of the program FIDISOL \cite{schoen}.
The equations are discretized on a non-equidistant
grid in $\bar{x}$ and  $\theta$.
Typical grids used have sizes $70 \times 30$, 
covering the integration region 
$0\leq\bar{x}\leq 1$ and $0\leq\theta\leq\pi/2$.

We construct numerically the black hole solutions for 
($k=1$, $n=1-8$), ($k=2$, $n=4-8$) and ($k=3$, $n=6-8$).
For $k=1$ and fixed $n$ a branch of black hole solutions emerges 
from the single globally regular solution,
when the horizon radius $x_{\rm H}$ is 
increased from zero \cite{kk}. This branch of black hole solutions
exists for arbitrarily large horizon size.
When $k \ge 2$ and $n \ge 2k$, however, 
a pair of globally regular solutions exists for a given set $(k,n)$.
Thus we find two branches of black hole 
solutions emerging from the corresponding globally regular solutions,
when the horizon radius is increased from zero.
Surprisingly, the two branches do not exist for arbitrarily large 
horizon size.
Instead they merge and end at a maximal value of the horizon size,
i.e.,~at the maximal value of the area parameter $x_{\Delta, max}$,
which depends on $k$ and $n$.

We show the ADM mass $\mu$ as a function of the area parameter $x_\Delta$
in Fig.~1. The mass increases monotonically with increasing $x_\Delta$.
For the new solutions with $k \ge 2$,
the maximal horizon size increases with increasing $n$. 
At the same time, the mass difference of the two regular solutions increases,
indicating a possible correlation between the soliton mass difference
and the maximal horizon size.

Considering the shape of the horizon, we observe that
the deviation from spherical symmetry is small, though.
We observe a ratio of circumferences for $k=1$ solutions 
of up to $L_e/L_p=0.988$,
and for $k=2$ and $k=3$ solutions of up to 
$L_e/L_p=0.984$ and $L_e/L_p=0.979$, respectively,
as seen in Fig.~2.
Thus the new black holes have a slightly more deformed horizon.

The inverse surface gravity $1/\kappa$ of the black hole solutions
is exhibited in Fig.~3.
For $k=1$ black hole solutions
the inverse surface gravity increases monotonically (except for $n=1$)
as a function of the area parameter $x_\Delta$.
For $k=2$ and $k=3$ black holes
it increases along the lower branch, and beyond the
transition to the upper branch it decreases back to zero.

Our numerical results indicate, that the mass relation (\ref{ash}),
obtained in the isolated horizon formalism, also holds for the new 
non-Abelian black hole solutions.
(In Fig.~1 for each branch the corresponding regular solution
is the reference point for the integration.)
Consequently, the mass of the regular solution on the upper branch is
related to the mass of the regular solution on the lower branch
via the horizon mass integral, performed along both branches.
A similar result was obtained previously for EYMH black holes
with dipole hair \cite{dipole}.

In \cite{iso2} the mass formula
$\mu_{\rm reg} = 1/2 \int_0^\infty(1-x_\Delta \hat{\kappa})d x_\Delta$
was derived and shown to hold for spherically symmetric EYM solutions in
\cite{eym_mass}.
We checked that the same formula also holds for the $k=1$ axially
symmetric solutions.

In Fig.~4 we illustrate the binding energy of the 
non-Abelian black hole solutions.
The binding energy is always calculated w.r.t.~the corresponding
regular solution.
Consequently, for $k=2$ and $k=3$ black holes the binding energy 
along the upper branches is smaller than the binding energy 
along the lower branches.
The difference in binding energy of the black hole solutions
for a given set of $(k,n)$ at the maximal value of
the area parameter represents the mass difference of the
corresponding pair of regular solutions.

\noindent
\parbox{\textwidth}{
\centerline{
{(1a)\epsfysize=5.0cm \epsffile{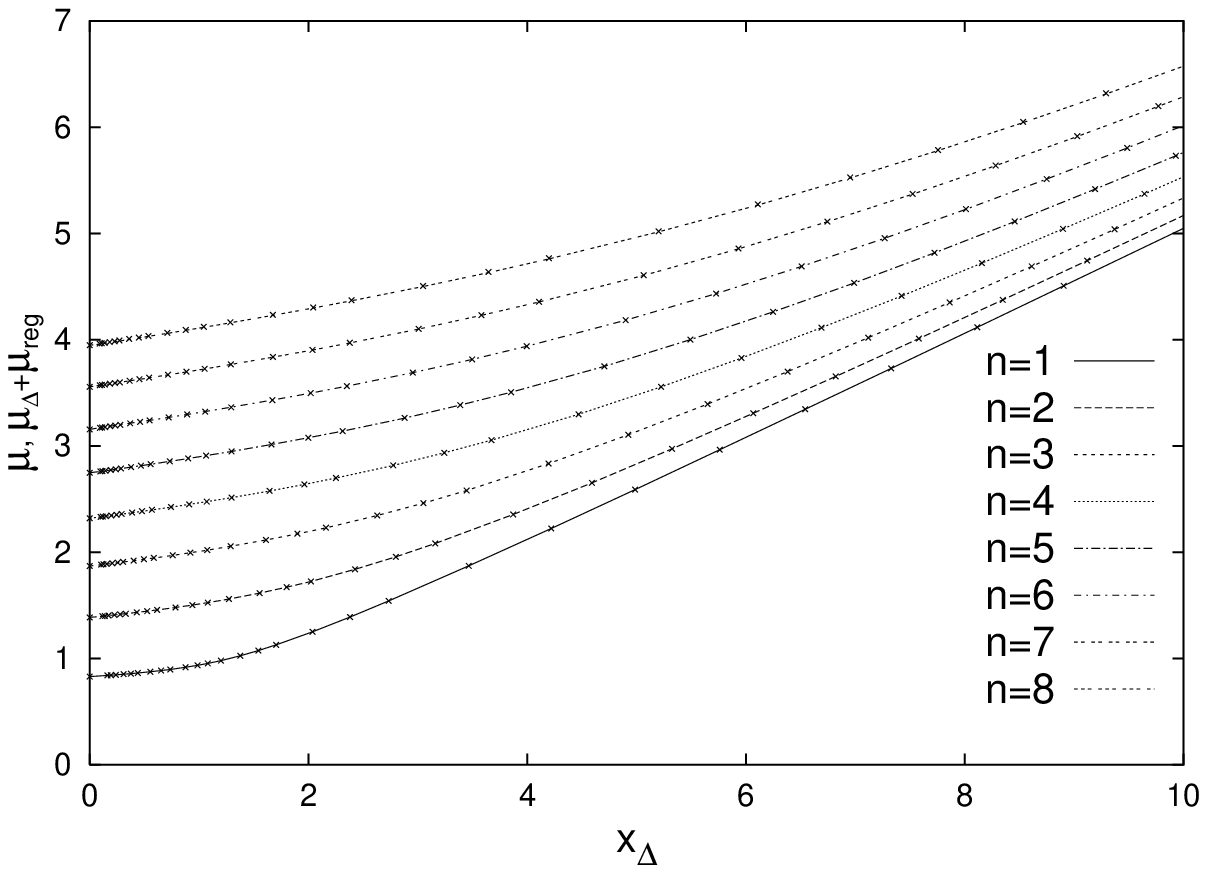} }
{(2a)\epsfysize=5.0cm \epsffile{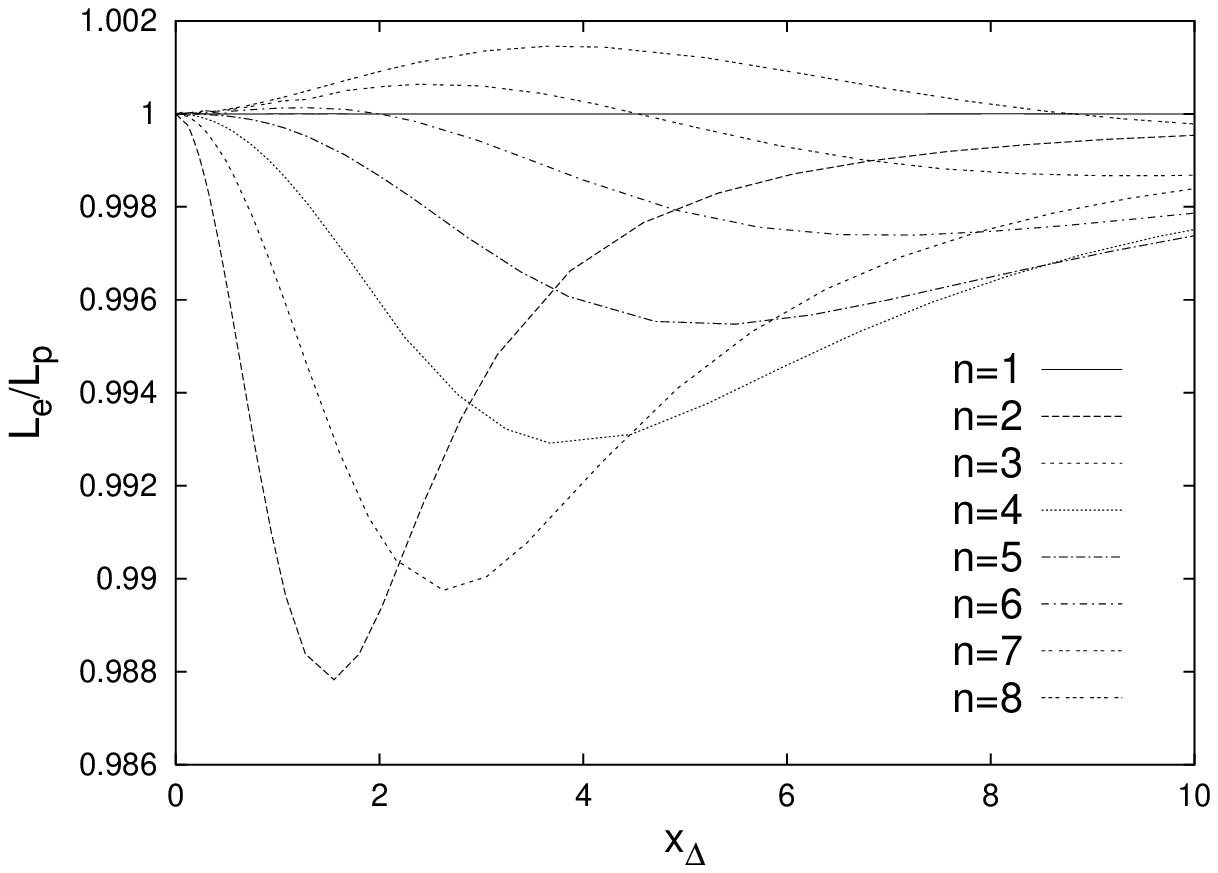} }
}\vspace{0.5cm}
\centerline{
{(1b)\epsfysize=5.0cm \epsffile{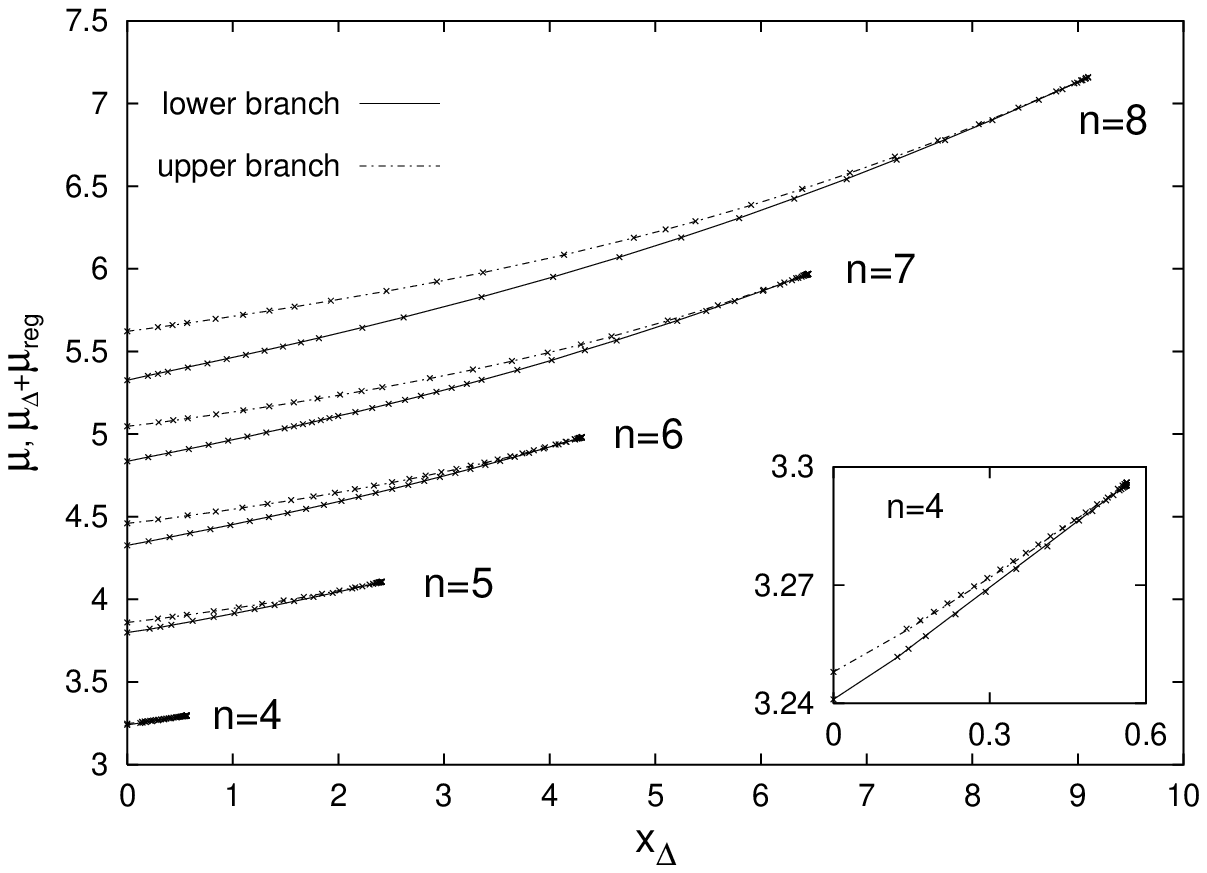} }
{(2b)\epsfysize=5.0cm \epsffile{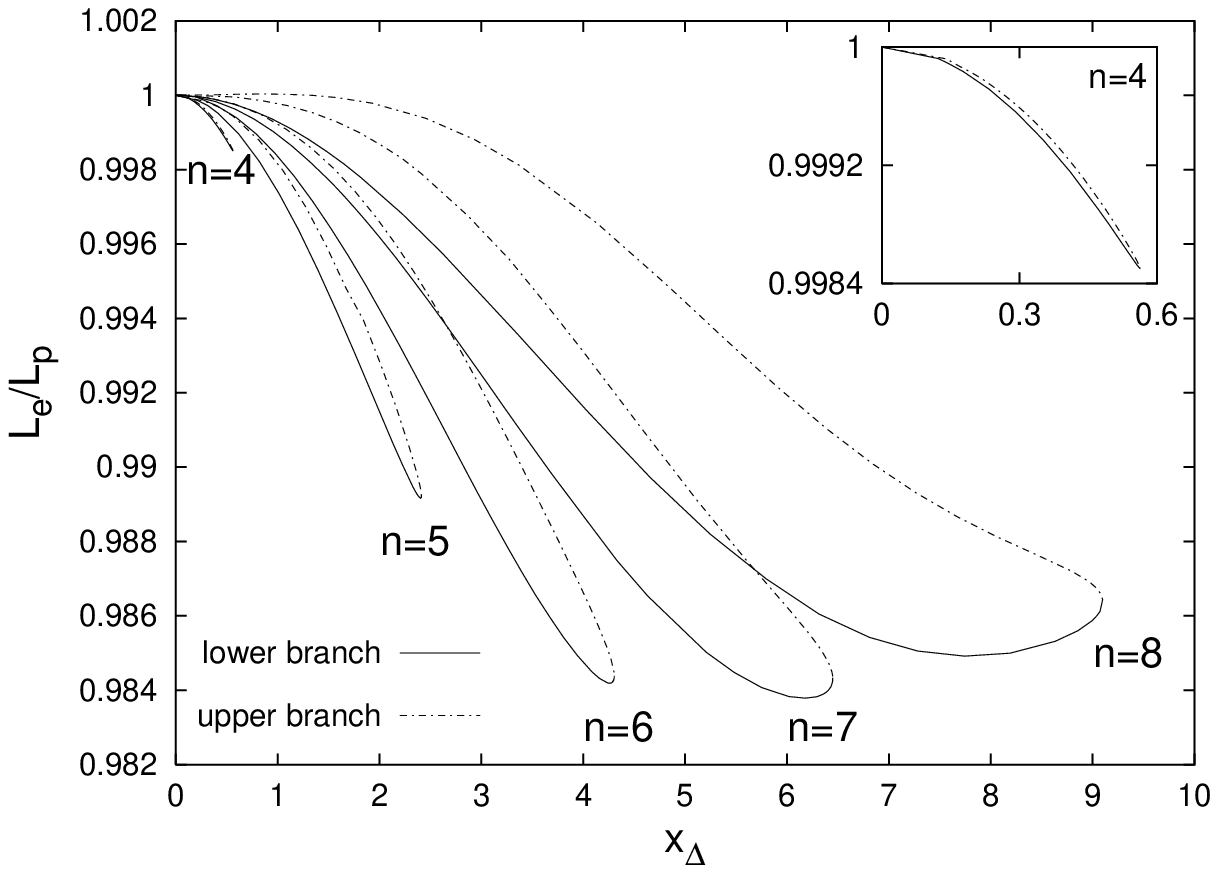} }
}\vspace{0.5cm}
\centerline{
{(1c)\epsfysize=5.0cm \epsffile{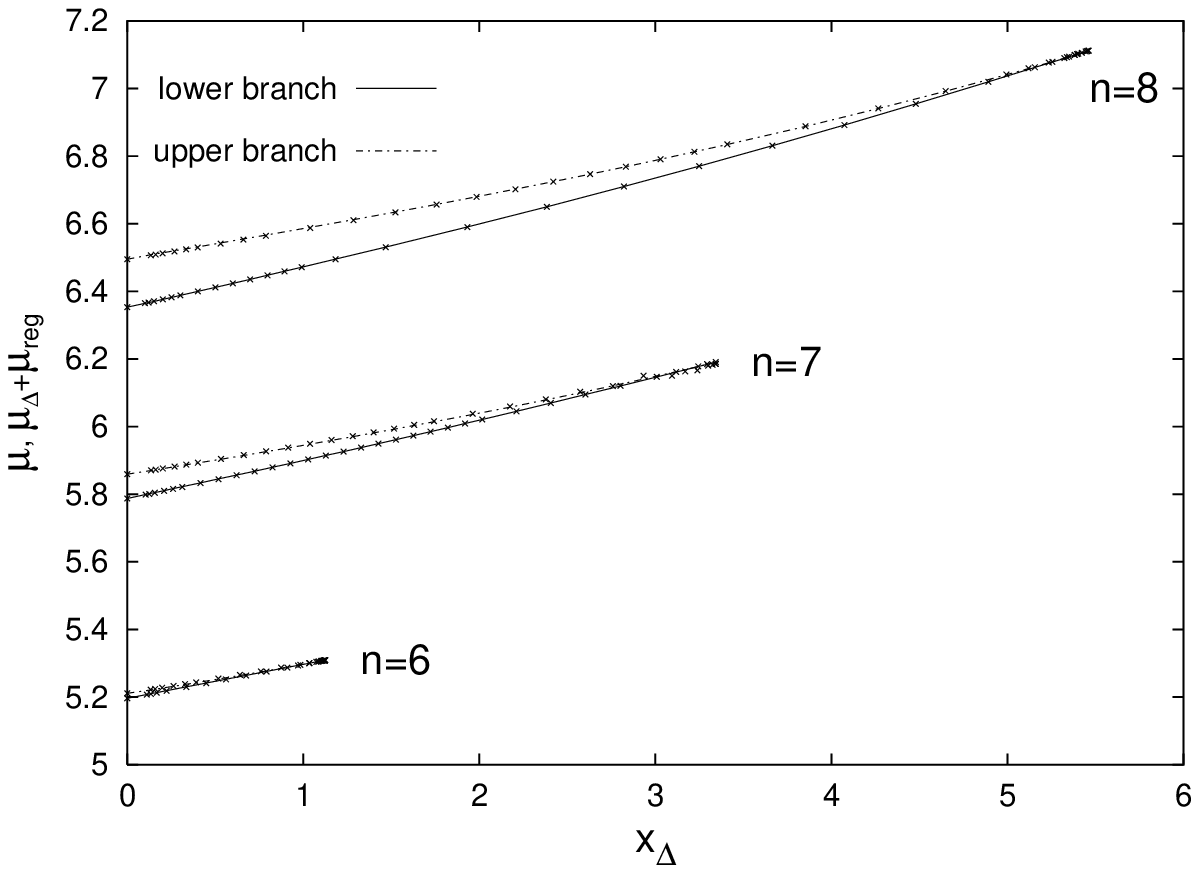} }
{(2c)\epsfysize=5.0cm \epsffile{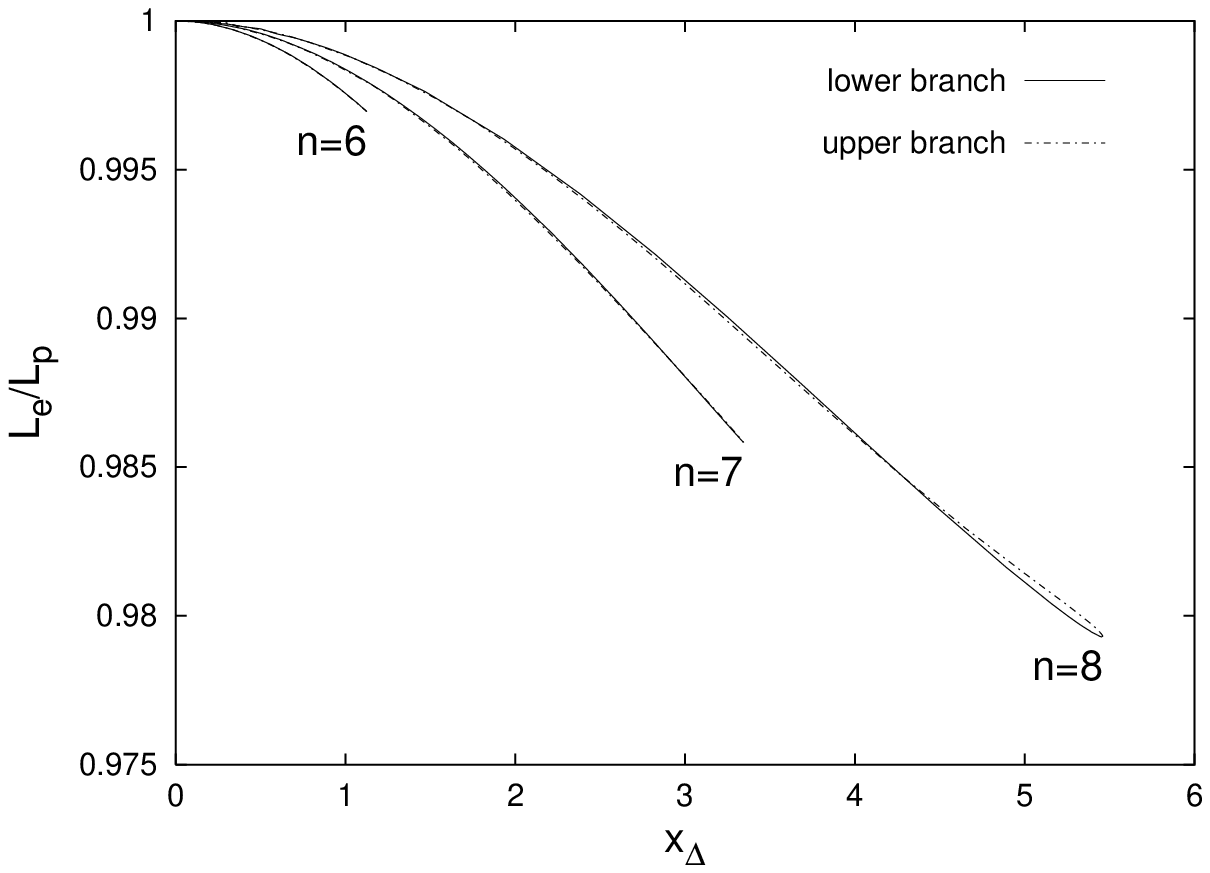} }
}\vspace{0.5cm}
{\bf Fig.~1} 
(a) The ADM mass $\mu$ is shown as function of the area parameter
$x_{\Delta}$ for $k=1$, $n=1-8$ (symbols $\times$). 
Also shown is $\mu_\Delta+\mu_{\rm reg}$ (lines).
(b) Same as (a) for $k=2$, $n=4-8$.
(c) Same as (a) for $k=3$, $n=6-8$.
{\bf Fig.~2}
Same as Fig.~1 for the ratio $L_e/L_p$. 
}\vspace{0.5cm}

\noindent
\parbox{\textwidth}{
\centerline{
{(3a)\epsfysize=5.0cm \epsffile{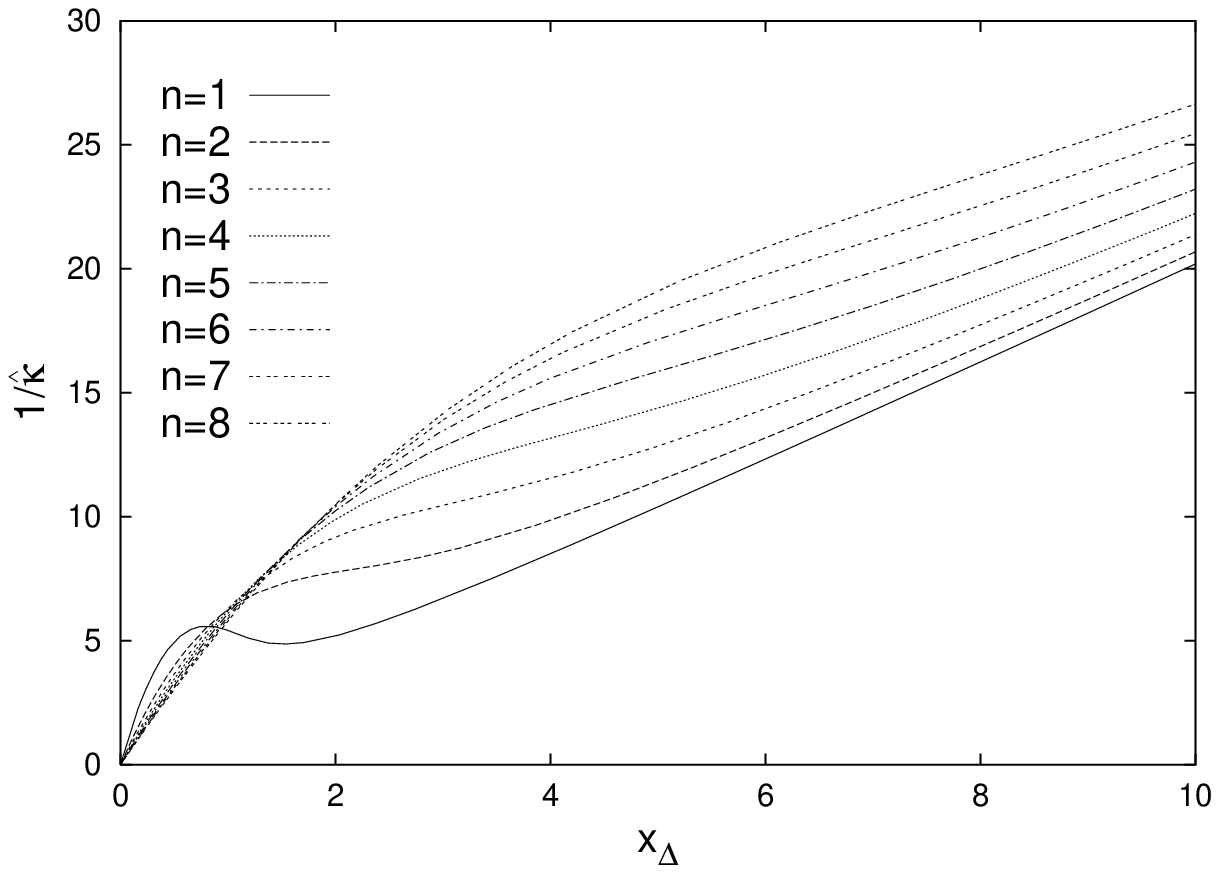} }
{(4a)\epsfysize=5.0cm \epsffile{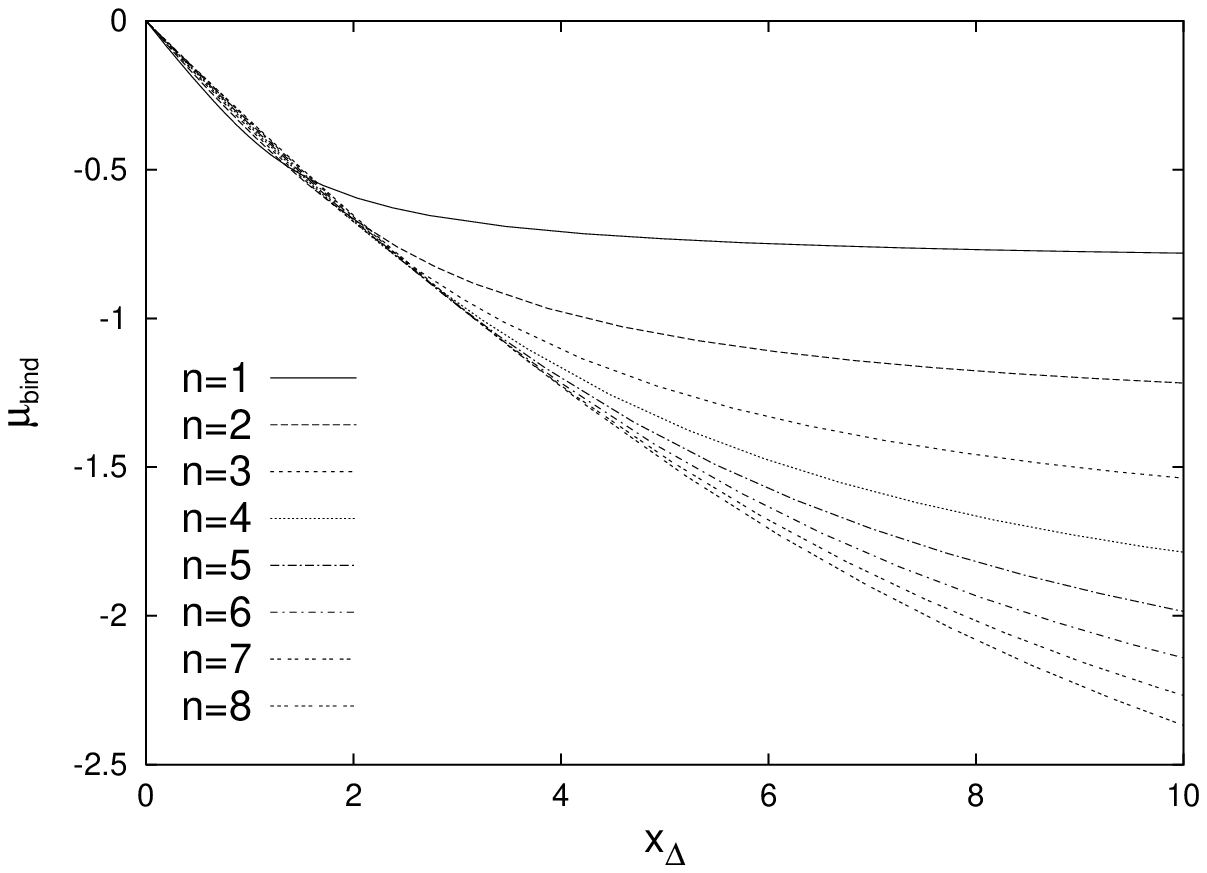} }
}\vspace{0.5cm}
\centerline{
{(3b)\epsfysize=5.0cm \epsffile{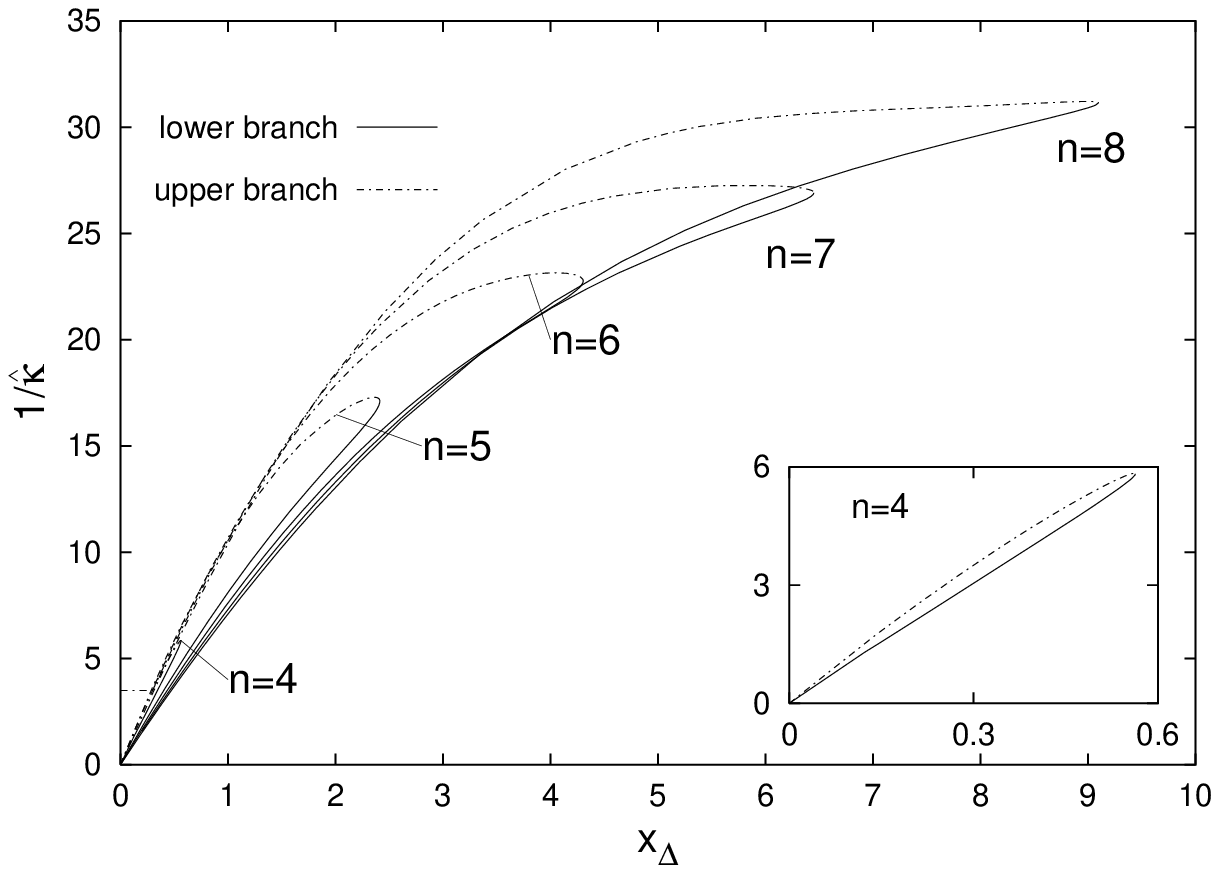} }
{(4b)\epsfysize=5.0cm \epsffile{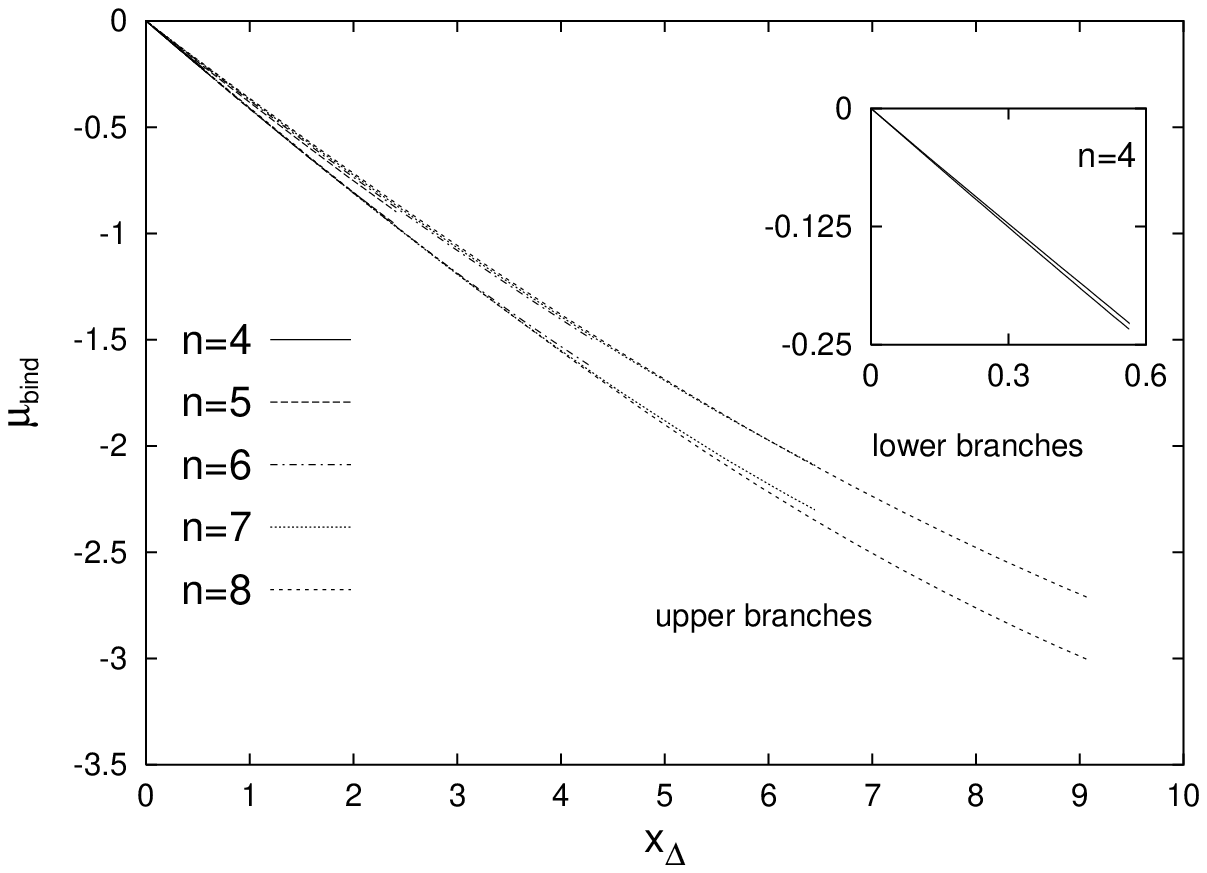} }
 }\vspace{0.5cm}
\centerline{
{(3c)\epsfysize=5.0cm \epsffile{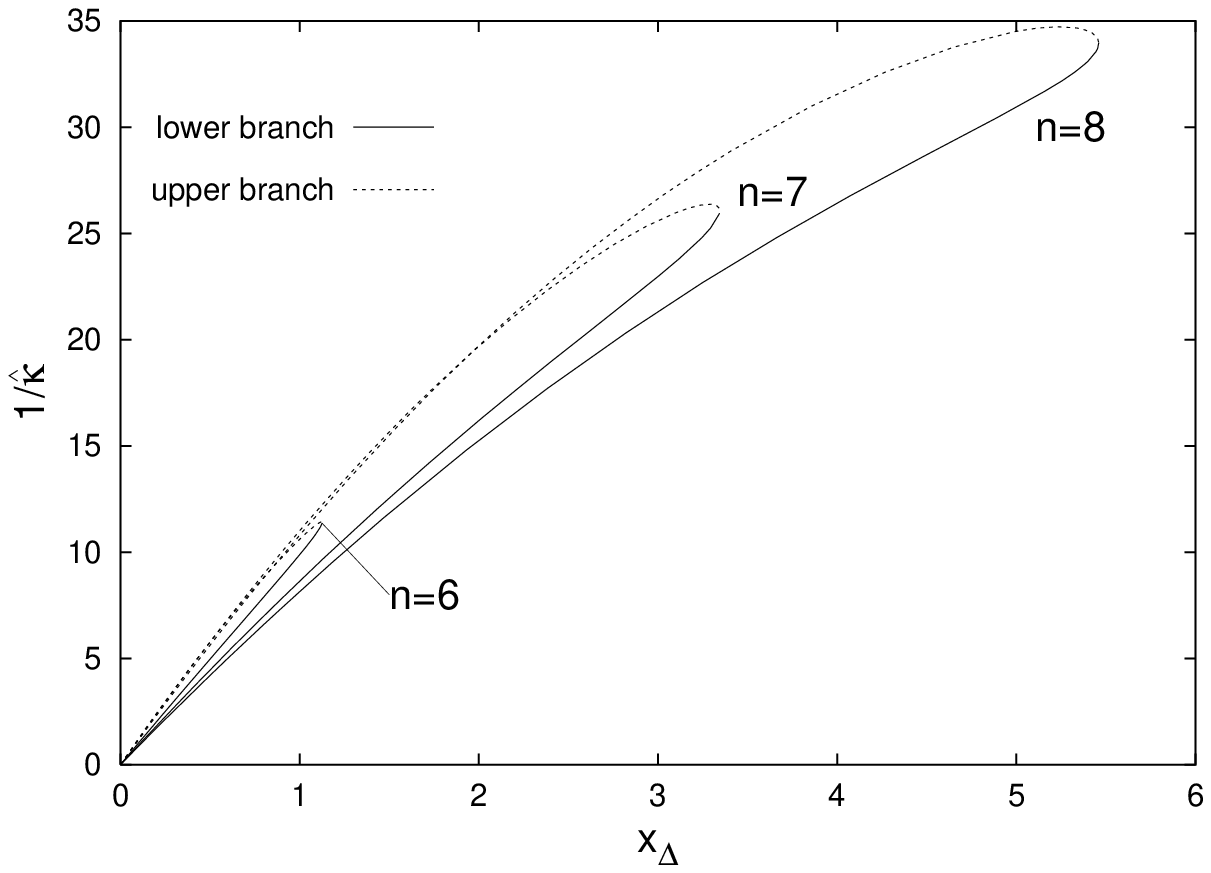} }
{(4c)\epsfysize=5.0cm \epsffile{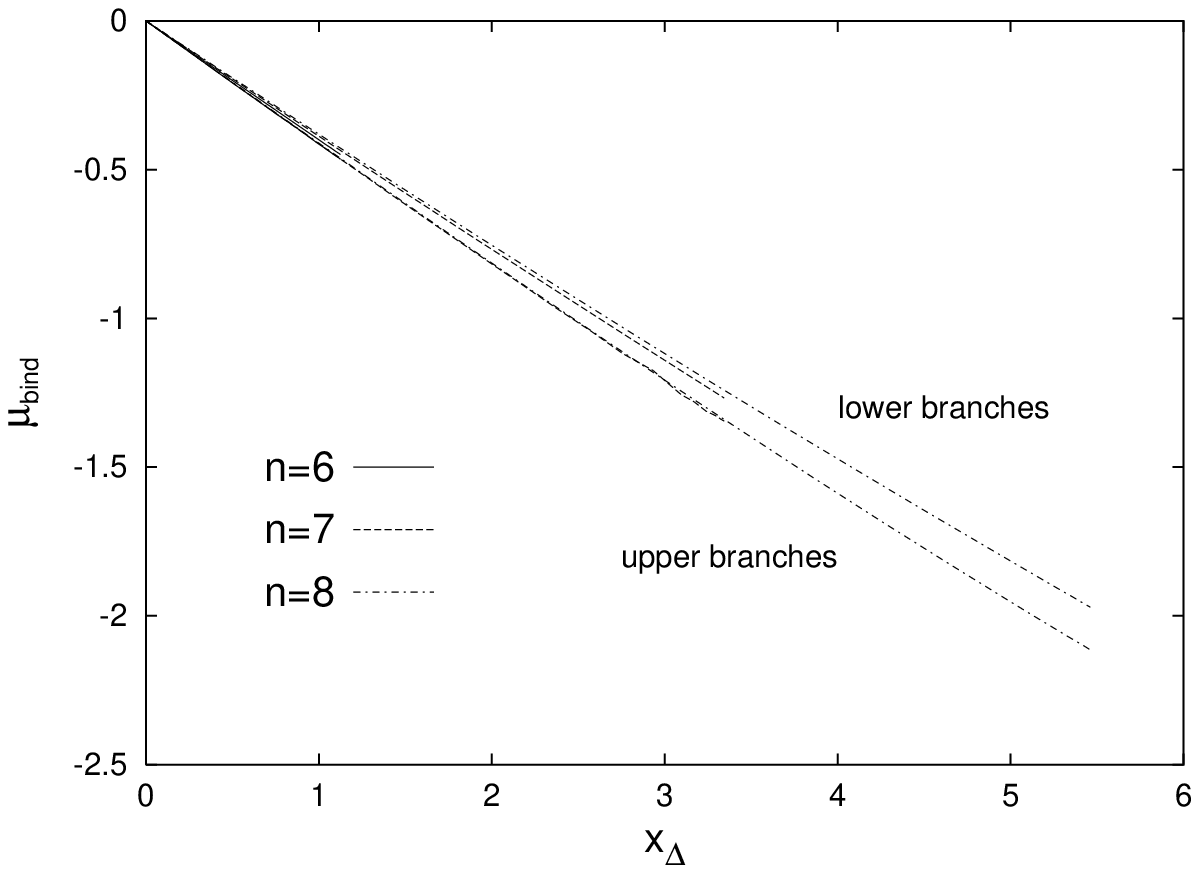} }
 }\vspace{0.5cm}
{\bf Fig.~3}
Same as Fig.~1 for the inverse surface gravity $1/\hat{\kappa}$.
{\bf Fig.~4} 
Same as Fig.~1 for the binding energy.
}\vspace{0.5cm}

\section{Conclusions}

We have constructed numerically new static axially symmetric
black hole solutions of EYM theory 
and investigated their properties.
These solutions are characterized by the set of integers $(k,n)$,
related to the polar and azimuthal angles, respectively.
In particular, we have obtained black hole solutions
for $k=2$, $n=4-8$, and $k=3$, $n=6-8$.

The $(1,n)$ solutions (with one node) exist most likely for any
integer $n\ge 1$ and for arbitrary horizon size.
The new $(2,n)$ and $(3,n)$ solutions have lower bounds on $n$, $n=2k$,
and upper bounds on the horizon size.
For each allowed set $(k,n)$
two branches of black hole solutions emerge from the two
globally regular solutions
and merge and end at the maximal value of the horizon size.
The existence of an upper bound of the horizon radius
is a surprising and interesting new feature for EYM solutions. Previously
a maximal value of the horizon size was known
only in theories with an in-built length scale already in Minkowski space,
hence it was conjectured, that ``there is no bound on the horizon radius
of hairy, static EYM black holes'' \cite{ashtekar}.
The underlying reason for the occurrence of this maximal horizon size 
for the new EYM black holes is yet to be understood.

We expect, that the $(2,n)$ and $(3,n)$ regular and black hole solutions
represent only the first sequences of new solutions,
and conjecture the existence of $(k,n)$ regular and black hole solutions
also for higher values of $k$.
The (sperically symmetric) $k=1$ black holes are unstable \cite{stab},
and there is all reason to believe, that the new $(k,n)$
solutions are unstable as well. 

Considering the new static axially symmetric solutions
from the isolated horizon formalism point of view,
we have verified the mass relation between the regular and the black hole
solutions, showing that the black hole mass is given by the sum of the mass
of the regular solution and the horizon mass.
In particular, the mass of the regular solution on the upper branch is
related to the mass of the regular solution on the lower branch
via the horizon mass integral, performed along both branches.
Interpreting the non-Abelian black holes as bound states of regular solutions
and Schwarzschild black holes \cite{iso2},
we have obtained the binding energy of these bound systems.

We note that the globally regular solutions were first obtained in EYMH
theory as limiting solutions for vanishing Higgs field \cite{kks4,ikks}.
Hence we expect that EYMH black holes can be found
from the new EYM black holes by gradually switching on the Higgs field.

Rotating EYM black hole solutions based on the static $k=1$ black hole solutions
are known \cite{rot}. It appears straightforward
to construct rotating EYM black hole solutions based on the static $k=2$
and $k=3$ black hole solutions.
In contrast, rotating regular EYM solutions do not appear to exist
\cite{radu},
though recently
the first rotating regular EYMH solutions have been constructed
\cite{radu,ikkn}. 

{\sl Acknowledgement}

R.I. gratefully acknowledges support by the Volkswagenstiftung,
and B.K. support by the DFG.

\vfill\eject

\end{document}